\documentclass[twocolumn,groupedaddress, superscriptaddress,preprintnumbers,amsmath,amsfonts,amssymb]{revtex4}
\usepackage{hyphenat}
\usepackage{float}
\usepackage[english]{babel}
\usepackage{lipsum}
\usepackage{microtype}
\usepackage[pdfstartview=FitH,linkcolor=blue,citecolor=red,urlcolor=green]{hyperref}
\usepackage{graphics,epsfig,graphicx}
\usepackage{dcolumn}
\usepackage{bm}
\usepackage[dvipsnames]{xcolor}
\usepackage[normalem]{ulem}
\usepackage{soul}

\usepackage{epstopdf}

\makeatletter
\def\maketitle{
\@author@finish
\title@column\titleblock@produce
\suppressfloats[t]}
\makeatother

\newcommand{\Ignore}[1]{}
\usepackage{epstopdf}
\begin{document}

\title  {Spin-dependent $\pi$$\pi^{\ast}$ gap in graphene on a magnetic substrate}

\author{P.\,M. Sheverdyaeva}
\email{polina.sheverdyaeva@ism.cnr.it}
\affiliation{Istituto di Struttura della Materia, Consiglio Nazionale delle Ricerche, Trieste, Italy}

\author{G.Bihlmayer}
\affiliation{Peter Grünberg Institut and Institute for Advanced Simulation, Forschungszentrum Jülich and JARA, 52425 Jülich, Germany}

\author{E. Cappelluti}
\affiliation{Istituto di Struttura della Materia, Consiglio Nazionale delle Ricerche, Trieste, Italy}

\author{D. Pacil\'{e}}
\affiliation{Dipartimento di Fisica, Universit$\grave{a}$ della Calabria, 87036 Arcavacata di Rende (CS), Italy}

\author{F. Mazzola}
\affiliation{Istituto Officina dei Materiali (IOM)-CNR, Laboratorio TASC, Strada Statale 14 km 163.5, 34149, Trieste, Italy}

\author{N. Atodiresei}
\affiliation{Peter Grünberg Institut and Institute for Advanced Simulation, Forschungszentrum Jülich and JARA, 52425 Jülich, Germany}

\author{M. Jugovac}
\affiliation{Elettra Sincrotrone Trieste, Strada Statale 14 km 163.5, 34149 Trieste, Italy}

\author{I. Grimaldi}
\affiliation{Dipartimento di Fisica, Universit$\grave{a}$ della Calabria, 87036 Arcavacata di Rende (CS), Italy}

\author{G. Contini}
\affiliation{Istituto di Struttura della Materia, Consiglio Nazionale delle Ricerche, 00133 Roma, Italy}

\author{A.\,K. Kundu\footnote{Present address: Condensed Matter Physics and Materials Science Division, Brookhaven Nat Lab, Upton, NY 11973, USA}}
\affiliation{Istituto di Struttura della Materia, Consiglio Nazionale delle Ricerche, Trieste, Italy}
\affiliation{International Center for Theoretical Physics (ICTP), Trieste, 34151, Italy}

\author{I. Vobornik}
\affiliation{Istituto Officina dei Materiali (IOM)-CNR, Laboratorio TASC, Strada Statale 14 km 163.5, 34149, Trieste, Italy}

\author{J. Fujii}
\affiliation{Istituto Officina dei Materiali (IOM)-CNR, Laboratorio TASC, Strada Statale 14 km 163.5, 34149, Trieste, Italy}

\author{P. Moras}
\affiliation{Istituto di Struttura della Materia, Consiglio Nazionale delle Ricerche, Trieste, Italy}

\author{C. Carbone}
\affiliation{Istituto di Struttura della Materia, Consiglio Nazionale delle Ricerche, Trieste, Italy}

\author{L. Ferrari}
\email{luisa.ferrari@ism.cnr.it}
\affiliation{Istituto di Struttura della Materia, Consiglio Nazionale delle Ricerche, 00133 Roma, Italy}

\date{\today}

\begin{abstract}

We present a detailed analysis of the electronic properties of graphene/Eu/Ni(111). By using angle and spin-resolved photoemission spectroscopy and ab initio calculations, we show that the Eu-intercalation of graphene/Ni(111) restores the nearly freestanding dispersion of the $\pi\pi^\ast$ Dirac cones at the K point with an additional lifting of the spin degeneracy due to the mixing of graphene and Eu states. The interaction with the magnetic substrate results in a large spin-dependent gap in the Dirac cones with a topological nature characterized by a large Berry curvature, and a spin-polarized van Hove singularity, whose closeness to the Fermi level gives rise to a polaronic band.

\end{abstract}

\maketitle

Graphene is a highly promising material for spintronics applications due to its large spin transport coherence length \cite{Tombros-07,Maassen-12}, and its particular band structure, especially for the massless Dirac cones and the nearly flat van Hove singularities (VHSs) \cite{VanHove1953,McChesney-10,Novoselov-05,Geim-07,Castro-09}. Despite graphene being a nonmagnetic material, it can acquire spin polarization by the proximity effect in contact with magnetic materials. However, on transition metals Fe, Co, and Ni the interaction with 3\textit{d} states strongly modifies the $\pi$$\pi^{\ast}$ states, which form complex hybrid bands and lose their linear character \cite{Pacile2014,Varykhalov-08,Varykhalov2009,Jugovac_2022}. One of the goals for spintronic would be turning graphene into a spin-conductor, finding supporting magnetic materials that preserve the graphene electronic properties almost unaltered \cite{Novoselov-05,Novoselov-12,Geim-07,Castro-09}. Removing the spin degeneracy may turn graphene into a spin field effect transistor \cite{Semenov2007}, spin valve \cite{Hill2006}, or spin-transfer torque device \cite{Zhou2010}. In turn, a spin-polarized VHS may lead to unconventional superconductivity \cite{Liu2014}, quantum phases\cite{Kai2011}, and insulating topological states \cite{Kang2020,Jugovac2023}. 

Recent works have addressed the behavior of rare earth metals intercalation in graphene/SiC \cite{Sung-17,Anderson-17,Daukiya-18,Watch-13,Link-19,Rosen-19} and graphene/ferromagnet heterostructures \cite{Voloshina-14,Huttmann-17,Jugovac2023}, where the {$\pi$} and $\pi^{\ast}$ bands maintain their characteristic linearity. At the same time, the heavy electron doping from the rare earth atoms brings the VHS to the proximity of the Fermi level (FL). 
A remarkable example is the Eu intercalated graphene/Ni(111) system, which has been studied by density functional theory (DFT) calculations and x-ray magnetic circular dichroism (XMCD) \cite{Huttmann-17}. The calculations show that the 2\textit{p} states of C atoms lose the characteristic hybridization with the 3\textit{d} states of Ni. The electron density of states (DOS) is spin-dependent and the XMCD data show that the Eu monolayer (ML) is ferromagnetic within the layer while being antiferromagnetically coupled to the Ni underlayer. The presence of the Ni film leads to a magnetic ordering above room temperature in the Eu layer, in contrast to the case of an iridium substrate that leads to a much lower transition temperature \cite{Schumacher-14}. This makes the graphene/Eu/Ni(111) system even more interesting for possible applications in spintronics. In this Letter, we investigate the electronic and magnetic properties of graphene/Eu/Ni(111), carried out by angle- and spin-resolved photoemission spectroscopy (ARPES and spin-ARPES), and DFT calculations. 
Our results show that, as a consequence of intercalation, the graphene is decoupled from the Ni states, and its states restore a linear dispersion in the proximity of the $\bar{\mathrm K}$ point. The interaction with the Eu 4\textit{f} states removes the spin degeneracy near the Dirac point and induces a large spin-dependent gap between the $\pi$$\pi^{\ast}$ bands. The large Eu-induced \textit{n}-doping leads to the emergence of a spin-polarized VHS at the FL and a pronounced quasiparticle band. In the vicinity of the energy gap, our calculations suggest the presence of a large Berry curvature, corroborating the topological nature of the bands \cite{Kane-2005,Flaschner-2016}. The latter, together with the time-reversal symmetry breaking, induced by the magnetic order, paves the way to the manifestation of exotic transport phenomena such as the quantum anomalous Hall effect.\par 

\begin{figure*}
\includegraphics[width =\textwidth]{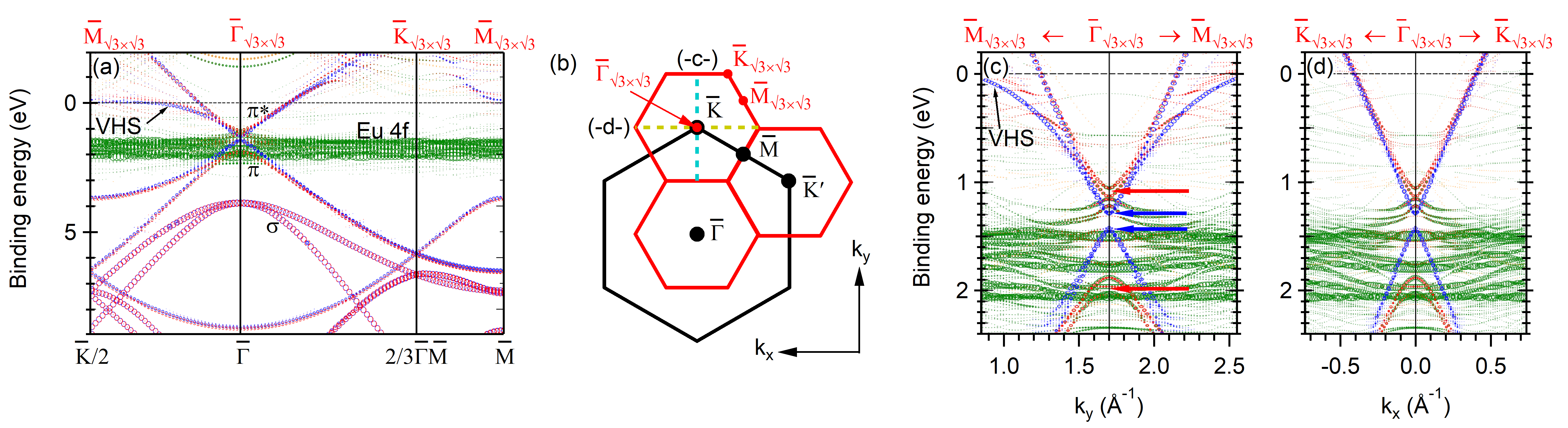}
\caption{\label{fig:DFT} (a) Spin resolved DFT band structure calculated for graphene/Eu/Ni/(111) along the main high symmetry directions. The red (blue) color indicates the graphene minority (majority) bands. The green (yellow) color indicates the Eu minority (majority) spin channels. The weight of the Ni states, in the background, is artificially reduced for clarity. (b) Schematic of the surface BZs of the ($1\times1$) graphene/Ni(111) (black), and the $(\sqrt{3}\times \sqrt{3})\ R30^\circ$ superstructure (red). The two high symmetry directions, investigated by theory and experiment, are drawn in yellow and light blue colors.
(c) and (d) Zoom of the calculated band along $\bar\Gamma$$\bar{\mathrm K}$ and p($\bar\Gamma$$\bar{\mathrm K}$), respectively. Red (blue) arrows indicate the minority (majority) cones.}.
\end{figure*} 

Fig.~\ref{fig:DFT}(a) shows the calculated band structure of graphene/Eu/Ni(111) system (for details see Sec.~I of Supplemental Material (SM)\cite{SupplInfo}). The system forms a $(\sqrt{3}\times\sqrt{3})\ R30^\circ$ superstructure with respect to the pristine graphene and Ni(111) lattices \cite{Huttmann-17} and the bands are folded as indicated in Fig.~\ref{fig:DFT}(b) and by the $x$-axis labels of Fig.~\ref{fig:DFT}(a). Because the superlattice-related replicas are not experimentally observed in our photoemission study as in others of similar systems \cite{Jugovac_2022,Jugovac2023}, we will not consider, in our discussion, the folding-induced bands, and for simplicity, will refer to the $(1\times1)$ surface for the indexing. The graphene’s $\pi\pi^\ast$ bands are shifted to higher binding energy with respect to the freestanding graphene (see Fig.~S1(a) of SM Sec.~I\cite{SupplInfo}), due to a combined effect of decoupling from the Ni substrate and electron doping from Eu. The position of the Dirac cone overlaps with the Eu 4\textit{f} minority states shown by green color (Fig.~\ref{fig:DFT}). The magnitude of doping (1.4 eV approximately) is similar to the one observed in other graphene/rare earth or alkaline systems \cite{Sung-17,Daukiya-18,Link-19,Rosen-19,McChesney-10}.

Figs.~\ref{fig:DFT}(c) and \ref{fig:DFT}(d) show a zoom of the region of the Dirac point along $\bar{\Gamma}$$\bar{\mathrm K}$ ($k_{y}$) and the perpendicular direction, p($\bar{\Gamma}$$\bar{\mathrm K}$) ($k_{x}$), respectively, indicated in the Fig.~\ref{fig:DFT}(b), revealing several important details in the graphene band structure. First, the $\pi\pi^\ast$ bands exhibit close to $\bar{\rm K}$ an almost linear dispersion typical of quasi-freestanding graphene \cite{Varykhalov-08,Pacile-13}, indicating that the Eu intercalation removes the direct graphene-Ni interaction. 
Second, the graphene states acquire a spin-polarization in some defined binding energy and wave vector regions where they interact with the Eu states. 
A large gap of about 900 meV opens for the spin-minority channel due to the sum of three effects: (i) the hybridization with the polarized 4\textit{f}-Eu states (about 750 meV), (ii) the intervalley coupling between $\bar{\mathrm K}$ and $\bar{\mathrm K'}$ due to the $(\sqrt{3}\times \sqrt{3})R30^\circ$ reconstruction (about 150 meV), and (iii) to a lesser extent the SOC (about 2 meV). For the majority-spin states, which are not affected by the interaction with the 4\textit{f}-Eu states, the gap is the effect of the last two sources and is about 150 meV. Moreover, the induced spin-polarization of the graphene bands leads to a 75 meV shift between the majority and minority bands. Furthermore, we notice 
in the $\bar\Gamma\bar{\rm K}$ direction
a strong bending of the Dirac $\pi^\ast$ band (Fig.~\ref{fig:DFT}(a)).
Such flat dispersion 
is related to the presence of the VHS
at the $\bar{\rm M}$ point, which
is shifted close to the FL
because of the large electron doping and
of the interaction with the substrate
(Fig.~S1(a) of SM Sec.~I\cite{SupplInfo}).
The interaction with the spin-polarized Eu \textit{sp}-bands
is expected to induce a spin polarization also for the
$\pi^\ast$ bands associated with the VHS.
This is more clearly seen in Fig.~S1(b) of SM\cite{SupplInfo}, which shows the calculated band structures of graphene on Eu without the Ni substrate, where the VHS shows two spin-split branches. On a Ni substrate, the indirect interaction with Ni states (see Fig.~S2 of SM Sec.~I\cite{SupplInfo}) leads to a stronger delocalization of the VHS's minority spin branch, so only the majority branch can be observed. Similar results were recently reported for the graphene/Eu/Co(0001) \cite{Jugovac2023}.

\begin{figure*}
\includegraphics[width =\textwidth]{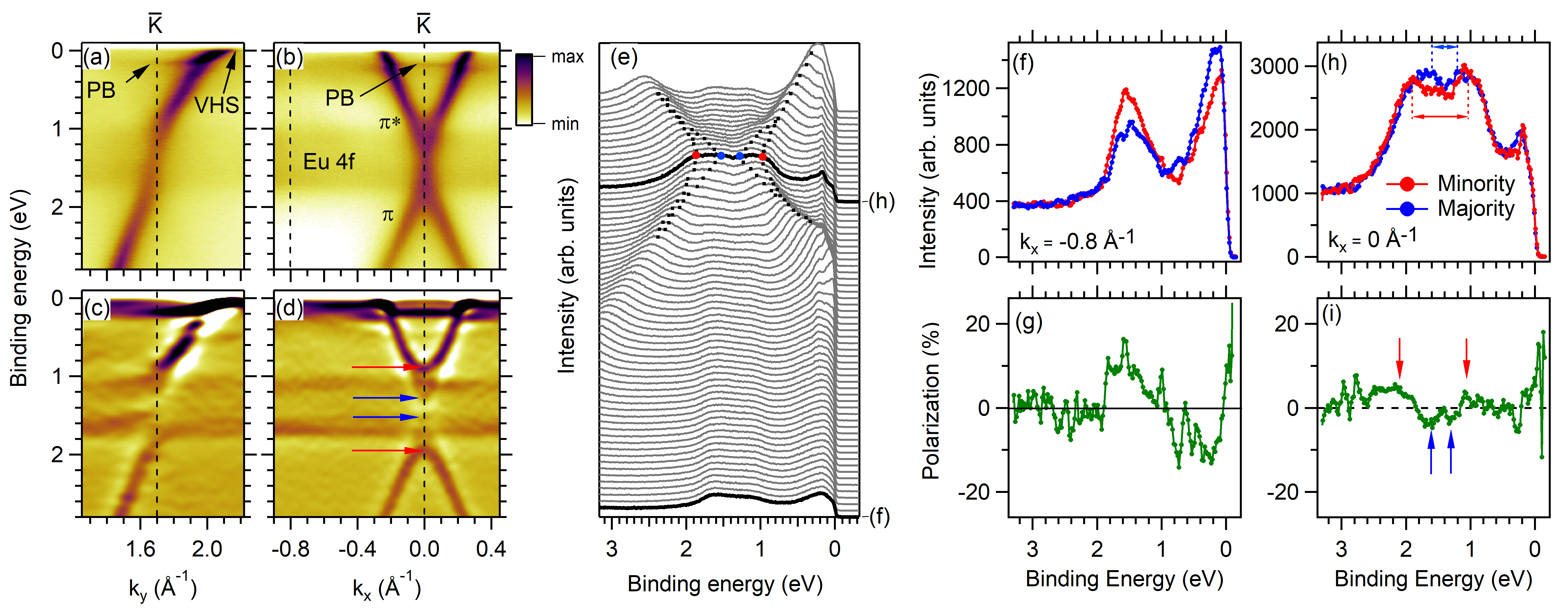}
\caption{\label{fig:arpes2}  
(a-d) Spin-integrated ARPES maps of graphene/Eu/Ni(111) along the (a,c) $\bar\Gamma\bar{\mathrm K}$ direction and (b,d) along the p($\bar\Gamma\bar{\mathrm K}$) directions, (a,b) as-taken and (c,d) second derivative along the energy axis. (e) EDC spectra extracted from the map (b), black dots follow the dispersion of the $\pi$ and $\pi^{\ast}$ states for the two spin channels. The bolded lines correspond to $k_{x} = 0$ and $k_{x} = -0.8$~\AA$^{-1}$. Red (blue) dots indicate the minority (majority) gap. The non-dispersive peak at 180 meV of binding energy is the PB band. (f,h) Minority-spin (red curves) and majority-spin (blue curves) EDCs at $k_{x} = -0.8$~\AA$^{-1}$ (f) and $k_{x} = 0$~\AA$^{-1}$ (h) (as the dashed lines indicate in (b)). Red (blue) arrows indicate the minority (majority) gap.
(g,i) Corresponding spin polarizations. Data were acquired at 25 K and a photon energy of 50 eV.}   
\end{figure*}

ARPES and spin-ARPES measurements were performed in order to confirm the dispersion and spin-polarization of the graphene $\pi$$\pi^\ast$  bands, in particular near the Dirac point, predicted by the theoretical calculations (for details see Sec.~II of SM\cite{SupplInfo}). 
Figs.~\ref{fig:arpes2}(a) and \ref{fig:arpes2}(b) show the ARPES data of the system close to $\bar{\rm K}$ along  $k_{y}$ and $k_{x}$, respectively, that can be compared to Figs.~\ref{fig:DFT}(c) and \ref{fig:DFT}(d). Dispersing graphene bands are crossed by a broad flat Eu 4\textit{f} state close to 1.5 eV. We can confirm the effects of doping and an almost linear dispersion of the $\pi$$\pi^\ast$ bands (see also Fig.~S4 of SM Sec.~II\cite{SupplInfo}), in good agreement with the theoretical predictions. A broadening of the $\pi$$\pi^\ast$ states in the proximity of the Eu states can be noticed. 
The second derivative spectra more clearly indicate the presence of two cone-like features and the opening of two different energy gaps at the $\bar{\rm K}$ point (Figs.~\ref{fig:arpes2}(c,d)), that amount to 250 $\pm$ 50 meV (blue arrows) and 1100 $\pm$ 50 meV (red arrows). Fig.~\ref{fig:arpes2}(e) reports the spin-integrated energy distribution curves (EDCs) extracted by the p($\bar\Gamma\bar{\rm K}$) map, where is possible to follow the dispersion of the $\pi$$\pi^{\ast}$ states for the two spin channels. These results are in agreement with the DFT predictions for two spin-dependent gaps, that we further confirm by spin-ARPES measurements. We analyzed the spin polarization at two different wave vectors: away from the Dirac point, where only the Eu 4\textit{f} states are visible ($k_{x} = -0.8$~\AA$^{-1}$), and at the Dirac point ($k_{x} = 0$~\AA$^{-1}$), where they overlap with the graphene states (black dotted lines in Fig.~\ref{fig:arpes2}(b)). Figs.~\ref{fig:arpes2}(f-i) show the corresponding majority and minority intensities (panels (f,h)) and spin polarization curves (panels (g,i)). Away from the $\bar{\rm K}$ point, we confirm the spin polarization of the Eu 4\textit{f} states: the majority- and minority- states of Eu (at $\sim{1.5}$ eV) show two similar large single-peak energy profiles with a non-complete spin polarization of about 15\%, due to the finite temperature and a non-complete magnetic saturation of the sample (Figs.~\ref{fig:arpes2}(f,g)). The sharp peaks just below the FL are the Ni states with prevalent polarization opposite to the Eu 4\textit{f} states. At the $\bar{\rm K}$ point, the majority- and minority-spin states show two-peaked profiles with clearly different energy distances between the peaks (Fig.~\ref{fig:arpes2}(h)) and we can distinguish the corresponding minima and maxima in the spin polarization spectra, better visible for $\pi$ states.
 We can estimate the peak separation about (400 $\pm$ 100) meV for the majority-spin channel and (1050 $\pm$ 100) meV for the minority-spin channel, in agreement with the majority and minority gap values provided by our theoretical results and spin-integrated ARPES data.

\begin{figure*}
\includegraphics[width=\textwidth]{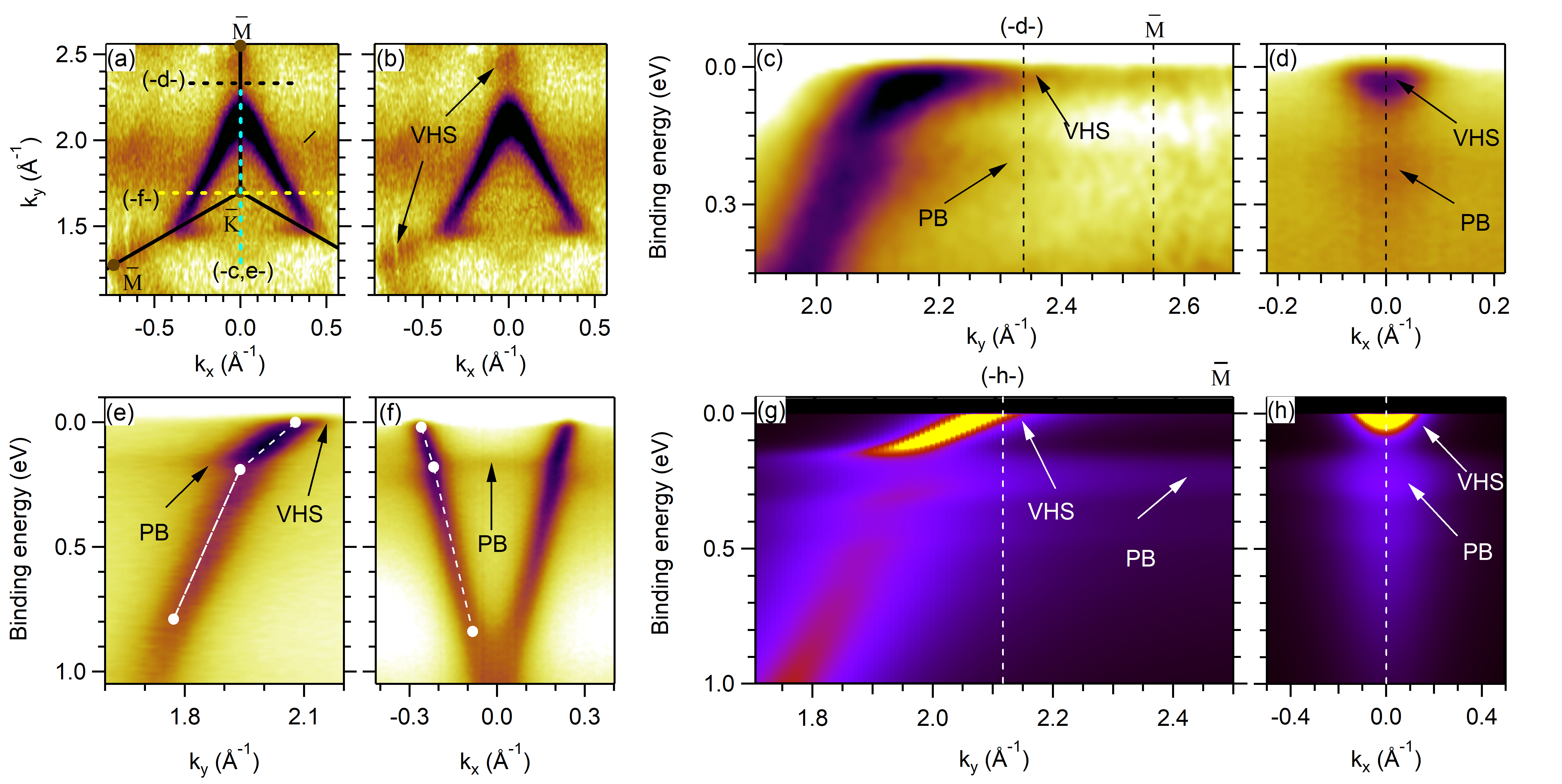}
\caption{\label{fig:polaron} (a,b) Experimental Fermi surface of graphene/Eu/Ni(111) with and without the surface Brillouin zone on top (h$\nu$=50 eV). (c) ARPES maps along $\bar{\rm K}\bar{\rm M}$ with an enhanced contrast (h$\nu$=40 eV). (d) Cut perpendicular to $\bar\Gamma\bar{\rm K}$ direction at the point indicated by the black dashed line in (a) and (c) (h$\nu$=75 eV); (e,f) ARPES spectra along $\bar\Gamma$$\bar{\rm K}$ and p($\bar\Gamma\bar{\rm K}$) showing the zoom on the vicinity of PB band, white dashed lines indicate the slope of the $\pi^\ast$ band in the proximity of the kink (h$\nu$=50 eV); (g,h) Theoretical simulations of the 
ARPES spectral function along the same cuts as in panels (c)-(d) including the many-body effects of the retarded electron-phonon self-energy. 
}

\end{figure*}

Close to the FL we can notice in Figs.~\ref{fig:arpes2}(a,c)
a strong bending of the $\pi^\ast$ band when $k_y$
moves towards the $\bar{\rm M}$ point.
Such feature
is in a very good agreement
with the DFT calculations that predict the VHS
of the $\pi^\ast$ band to be located
just above the FL.
As a consequence, we are able to detect it only partially, similarly to other highly \textit{n}-doped graphene systems \cite{Jugovac2023,McChesney-10,Ehlen2020}.
Figs.~\ref{fig:polaron}(a) and \ref{fig:polaron}(b) show the experimental Fermi surface where the VHS is marked by black arrows. In the ARPES maps
(Figs.~\ref{fig:polaron}(c,d)) it can be observed as a weak intensity right at the FL that extends towards the $\bar{\rm M}$ point.
Due to the low intensity of VHS, we were not able to access experimentally its spin polarization. The agreement between the DFT calculations and our ARPES and spin-ARPES data for the Dirac cone suggests that the VHS preserves a spin-polarization with minority character, similarly to the graphene/Eu/Co system\cite{Jugovac2023}. 

In Figs.~\ref{fig:arpes2}(a-e) we can also notice
a second low-dispersive feature, which stems out from the Dirac cone at about 180 meV of binding energy. We identify the nature of this spectral feature as related to polaronic effects, and we denote it as
polaronic band (PB). Such spectral feature, which is also
clearly visible in Figs.~\ref{fig:polaron}(c-f),
is absent in the DFT simulations. We can observe that it extends towards $\bar{\rm M}$ as a replica of the VHS in Fig.~\ref{fig:polaron}(c). A kink at a similar binding energy is commonly observed in graphene bands and is attributed to el-ph coupling \cite{Bostwick-07N, Papagno-12, Fedorov-14, Haberer-13}. While a band similar to PB is observed on highly \textit{n}-doped graphenes close to $\bar{\rm M}$, its reports at $\bar{\rm K}$ are quite few \cite{Link-19, Jugovac2023, McChesney-10, Ichinokura2023, Rosen2020}, and are often attributed to the substrate or dopants \cite{Jugovac2023,Ichinokura2023,Rosen2020}. This hypothesis is in contrast with the variety of systems where this band was observed. The 180 meV binding energy corresponds to the phonon mode of graphene, and indeed several studies associated the PB band to a 
strong el-ph coupling in the polaronic regime \cite{Link-19,McChesney-10}, boosted by the closeness of the VHS.
Our findings support this view, and it is evident
from Fig.~\ref{fig:polaron}(d) where the main spectral feature at low-energy $\omega\approx 0$ associated with the VHS, is accompanied by a weak replica at $\omega \approx 160-180$ meV, where no DFT bands are expected, with the typical profile of polaronic replica \cite{Link-19,Riley2018}.  We confirm the robustness of this view by performing many-body self-consistent calculations of the retarded el-ph self-energy in the strong-coupling regime (see Sec.~III of SM\cite{SupplInfo} for details).
A standard tight-binding model was employed with nearest-neighbor hopping $t_0=1.92$ eV chosen to reproduce the experimental values of the Dirac velocity,
 and the FL was set to reproduce the
 experimental Fermi surface.
We simulated the anisotropy of the el-ph
coupling induced by the VHS, obeying the three-fold symmetry,
by using different coupling constants $\lambda$ along the different directions.
From the ratio of the low-energy and high-energy slopes
(see Figs.~\ref{fig:polaron}(e,f)) we estimate
$\lambda=1.63$ and $\lambda=0.3$ along $\bar{\rm K}\bar{\rm M}$
and $p\bar{\rm \Gamma}\bar{\rm KM}$, respectively.
A constant quasi-particle damping $\gamma_{\rm imp}$ = 80 meV
was included to account for disorder/impurity scattering.
The resulting spectral features along $\bar\Gamma\bar{\rm K}$, and perpendicular to it (see dashed lines), are shown in Figs.~\ref{fig:polaron}(g,h). The so-extended observed polaronic band is well reproduced when the VHS is placed in the proximity of the FL. The result of the model, using the same strength of el-ph coupling but with the VHS far from the FL, leads to a PB band less strong and evident (see Fig.~S5 and Sec.~III of SM~\cite{SupplInfo} for more information). This demonstrates a direct relation of PB to the VHS band at the FL and hence to the large \textit{n}-doping induced by Eu. 

\begin{figure}
\includegraphics[width =0.98\columnwidth]{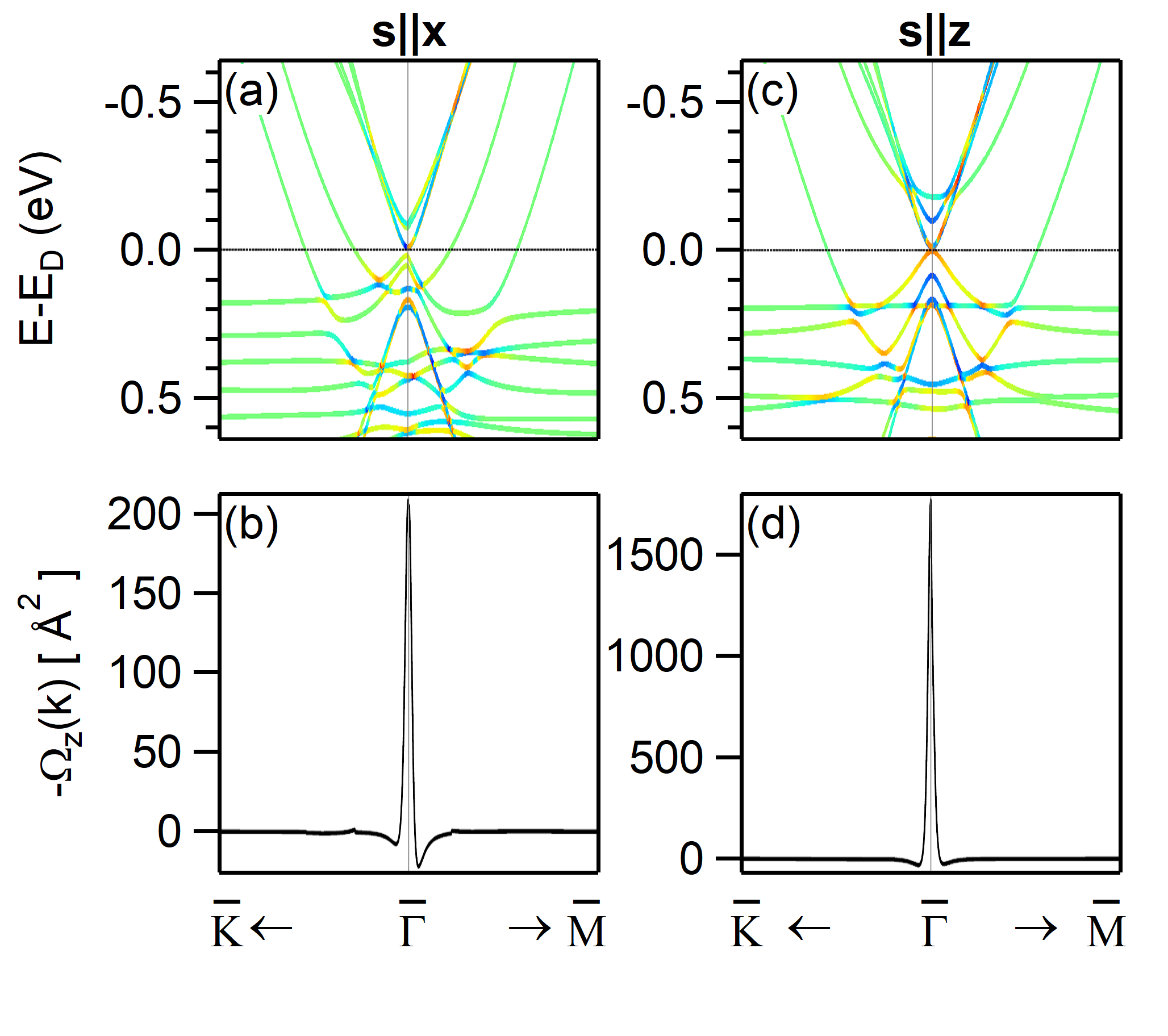}
\caption{\label{fig:berry}  Berry curvature calculation. (a) \textit{k}-resolved Berry curvature for the graphene/Eu system with in-plane spin direction in the proximity of the Dirac cone along $\bar{\mathrm K}$$\bar{\mathrm \Gamma}$$\bar{\mathrm M}$ direction. The binding energy is referred to the Dirac point. The color scale is logarithmic where red=positive, green=zero, blue=negative.
(b) Band-integrated \textit{k}-resolved Berry curvature, taken on the valence states. (c,d) same as (a,b) for an out of plane spin direction. }
\end{figure}

The above reported large and spin-dependent graphene's gap can give rise to a number of effects for spintronic applications, such as spin-filtering, spin-selective injections or the quantum anomalous Hall effect. Regarding the latter, we evaluated the topological character of the gap by calculating the Berry curvature \cite{Rybkin2022,Zanolli-2018} of a $(\sqrt{3}\times\sqrt{3})R30^\circ$ Eu in a contact with a freestanding graphene (see Fig.~S1(b) of SM Sec.~I\cite{SupplInfo}). 
Fig.~\ref{fig:berry}(a) shows the Berry curvature  $\Omega^z$${_n}$ \cite{Qiao-2018} as a function of binding energy and wave vector in the proximity of the Dirac cone, for the in-plane spin polarization, corresponding to the present experimental case. 
We can observe a finite Berry curvature that gives rise to a sizeable $k$-resolved integral curvature (Fig.~\ref{fig:berry}(b)). 
Higher values can be obtained for the out-of-plane spin polarization direction (Figs.~\ref{fig:berry}(c,d)) that can be experimentally realized, for example, on a thin Co substrate \cite{Rougemaille2012}. Due to the $(\sqrt{3}\times\sqrt{3})R30^\circ$ reconstruction, there is an overlap of the Dirac cone with the Eu \textit{sp}-states, and a superposition of two opposite contributions from $\bar{\rm K}$ and $\bar{\rm K'}$ leading to a lowering of the integral curvature. In order to cross-check
our findings, we also analyzed the ($2\times2$) graphene/Eu system (Fig.~S6(a) of SM Sec.~IV\cite{SupplInfo}),  that can be realized on Ir(111) \cite{Schumacher-14}. The disentanglement of the contributions from $\bar{\rm K}$, $\bar{\rm K'}$ and Eu \textit{sp}-band provides a simpler electronic pattern in the proximity of the Dirac point, and leads to an increase of the Berry curvature values by several orders of magnitude (see Figs.~S6(b,c) of SM Sec.~IV\cite{SupplInfo}). 
The development of a finite Berry curvature to across the energy gaps corroborates the topological nature with a non-zero topological index of the bands \cite{Xiao-2010} which gives rise to such energy separation. This finding elucidates the possibility of turning the graphene-Eu heterostructures into candidates for quantum anomalous Hall effect devices. 

In conclusion, through theoretical studies and experimental evidence, we show that graphene on a magnetic Eu presents an almost inaltered dispersion of the $\pi\pi^\ast$ states together with a lifting of spin degeneracy and a large electron doping. DFT predict that, at the Dirac point, a large energy gap opens for the minority spin channel due to the spin-dependent interaction with Eu 4$\textit{f}$ states, while in the opposite spin channel, which is undisturbed by substrate states, a smaller gap opens. ARPES and spin-ARPES data provide experimental evidence for this spin-polarized gap. We furthermore report for a spectroscopic signature of a polaronic band and demonstrate its direct relation to the large Eu-induced \textit{n}-doping and the VHS. A finite Berry curvature across the gap and the breaking of the time-reversal symmetry allow for the emergence of a quantum anomalous Hall effect in this system.

\section*{Acknowledgements}
G.B. and N.A. gratefully acknowledge the computing time granted by the JARA Vergabegremium and provided on the JARA Partition part of the
supercomputer JURECA at Forschungszentrum Jülich. N.A. acknowledges DFG support within CRC1238, project no. 277146874-CRC 1238 (subproject C01)".
E.C. acknowledges financial support from PNRR MUR Project No. PE0000023-NQSTI.
Authors acknowledge Elettra Sincrotrone Trieste for providing access to its synchrotron radiation facilities and for financial support under the SUI internal project and EUROFEL-ROADMAP ESFRI of the Italian Ministry of Education, University, and Research. This work has been partly performed in the framework of the nanoscience foundry and fine analysis (NFFA-MUR Italy Progetti Internazionali) facility.

%
%

\clearpage
 \setcounter{figure}{0}
 
\title{SUPPLEMENTAL MATERIAL for:
Spin-dependent $\pi$$\pi^{\ast}$ gap in graphene on a magnetic substrate}
\maketitle
\onecolumngrid
\clearpage
\renewcommand{\thefigure}{S\arabic{figure}}%

\section{Details of the density functional calculations}

Density functional theory (DFT) calculations including spin-orbit coupling (SOC) were carried out using the generalized gradient approximation \cite{Perdew-96}. To properly account for the van der Waals (vdW) interactions, the structural relaxation was performed employing the vdW-DF2 \cite{Lee-2010} with a revised Becke (B86b) exchange energy functional \cite{Becke-1986,Hamada-2014}. The full potential linearized augmented plane wave method, as implemented in the Fleur code, was employed \cite{wortmann_daniel_2023_7778444}. For the data shown in Fig.~1 of the main text, a relaxed structure with nine layers of Ni(111) was used as substrate. The Eu layer density was set to one atom per $(\sqrt{3}\times\sqrt{3})\ R30^\circ$ Ni(111)/graphene cell. A Hubbard U correction with U=6.7 eV and J=0.7 eV was used to capture the localized nature of the Eu 4\textit{f} states. We chose an out-of-plane spin-quantization axis to keep the three-fold symmetry, as an in-plane SOC would lead to breaking of the hexagonal symmetry of the Brillouin zone, non observed in our ARPES data, in analogy to Ref.~\cite{Jugovac2023}. The Ni magnetization defines the spin-quantization axis for majority and minority spin. Eu is divalent, with a fully spin-polarized 4\textit{f}$^7$ configuration. The Eu 4\textit{f} states at about 1.5 eV below the FL are of minority character because of the antiferromagnetic coupling between Ni and Eu \cite{Huttmann-17}, whereas the majority channel of the 4$\textit{f}$ states lies at around 9 eV above the FL. For the freestanding $(\sqrt{3}\times\sqrt{3})\ R30^\circ$ and $2\times2$ graphene/Eu systems the U values were set to 4.2 and 2.7 eV, correspondingly, in order to shift the Eu 4\textit{f} states to the graphene’s Dirac point.

\begin{figure}[h]
\includegraphics[width=0.99\textwidth]{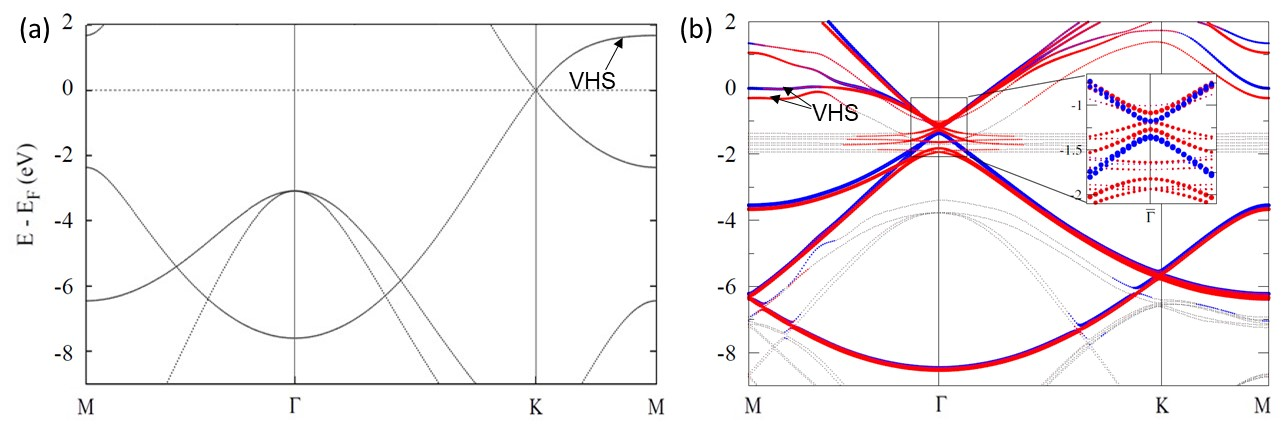}
\caption{\label{fig:DFT-Eu}  (a) Freestanding graphene band structure DFT calculation (general gradient approximation, a$_0$=2.47\AA). (b) DFT band-structure calculation for graphene/Eu along the $\bar\Gamma$-$\bar{\mathrm M}$ and $\bar\Gamma$-$\bar{\mathrm K}$-$\bar{\mathrm M}$  directions of the reduced Brillouin zone, U$_f$=4.2 eV. The arrows mark the polarized VHS band of graphene. 
}
\end{figure}
\begin{figure}[h]   
\includegraphics[width=0.55\textwidth]{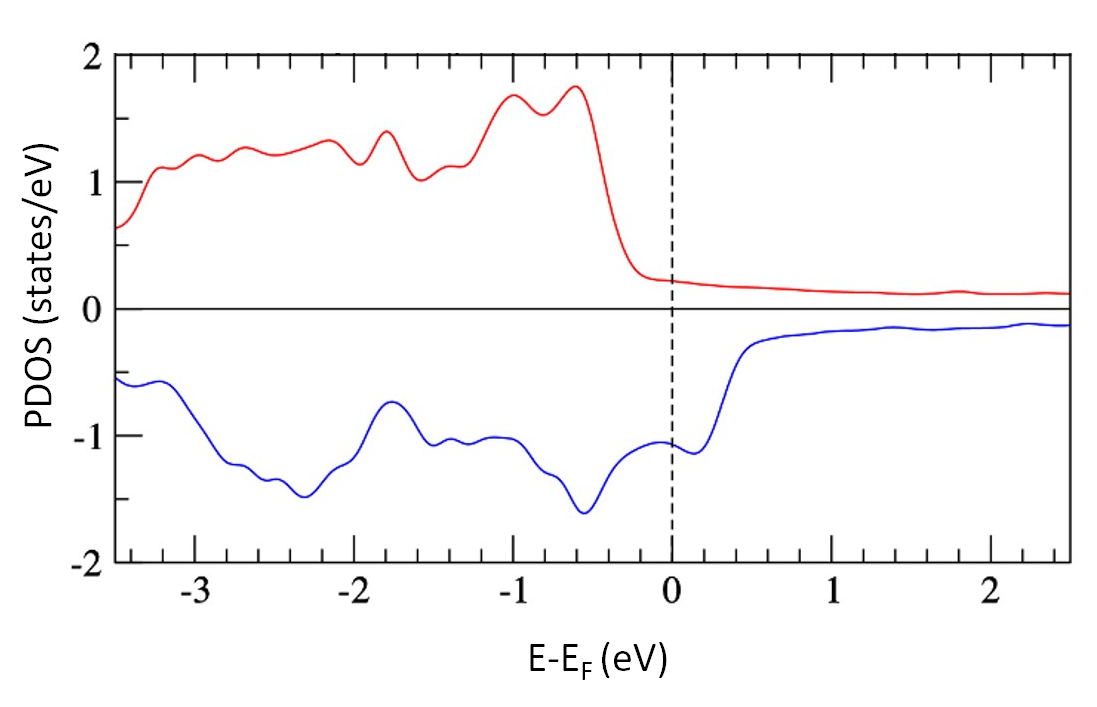}
\caption{\label{fig:PDOS} Spin-polarized local projected density of states  (SP-PDOS) calculated for surface 3\textit{d} orbitals of a Ni below Eu in graphene/Eu/Ni/(111) system. The Ni states close to the Fermi level are mainly of majority character, therefore they hybridize with Eu and graphene's minority  states.
}
\end{figure} 
\clearpage

\section{Details on the experimental characterization and additional figures}

The graphene/Eu/Ni(111) sample was prepared \textit{in situ}. A Ni(111) film with a thickness of about 4 nm was grown at room temperature on a W(110) single crystal. Graphene was grown by chemical vapour deposition (CVD) of ethylene on top of the Ni film held at 725 K. The Eu was evaporated on graphene from a Knudsen cell keeping the sample at 700 K in order to produce directly an intercalated layer without oxygen contamination. The low energy electron diffraction (LEED) showed the $(\sqrt{3}\times\sqrt{3})\ R30^\circ$ pattern with respect to the Ni(111) and graphene/Ni systems at the saturation coverage of Eu (Fig.~S3). High-resolution angle-resolved photoemission spectroscopy (ARPES) and spin-ARPES data were collected at the VUV-Photoemission and APE-LE beamlines of the Elettra Synchrotron, with  horizontal linear polarized light and a photon energy range of (35-80) eV, at the temperature of 25 K. The sample was remanently magnetized along the easy axis ($\bar\Gamma\bar{\mathrm K}$ direction of the Ni film matching with the [001] direction of the W(110) substrate). The VLEED polarimeter was aligned with the analyser slit, probing in-plane spin polarization. During the measurements, the easy axis was aligned at 30$^\circ$ with respect to the polarimeter. The majority and minority intensities were extracted using the Sherman function value of 0.25. The energy and angular resolution for ARPES measurements were 10 meV and $0.1^\circ$ respectively, while, for spin-ARPES measurements, the energy and angular resolution were about 90 meV, and 0.5$^\circ$ respectively.
  

\begin{figure}[h]
\includegraphics[width =0.40\textwidth]{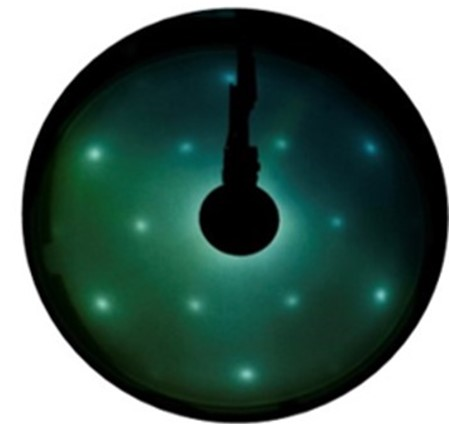}
\caption{\label{fig:LEED} 
LEED pattern of the intercalated system obtained at E = 100 eV. The LEED pattern indicates a homogenous and well-developed surface. It is the overlapping of the $(1\times1)$ graphene spots (outermost) and the $(\sqrt{3}\times \sqrt{3})R30^\circ$ superstructure spots (inner ones), as expected.
}
\end{figure} 
\clearpage

\begin{figure}[h]
\includegraphics[width =0.60\textwidth]{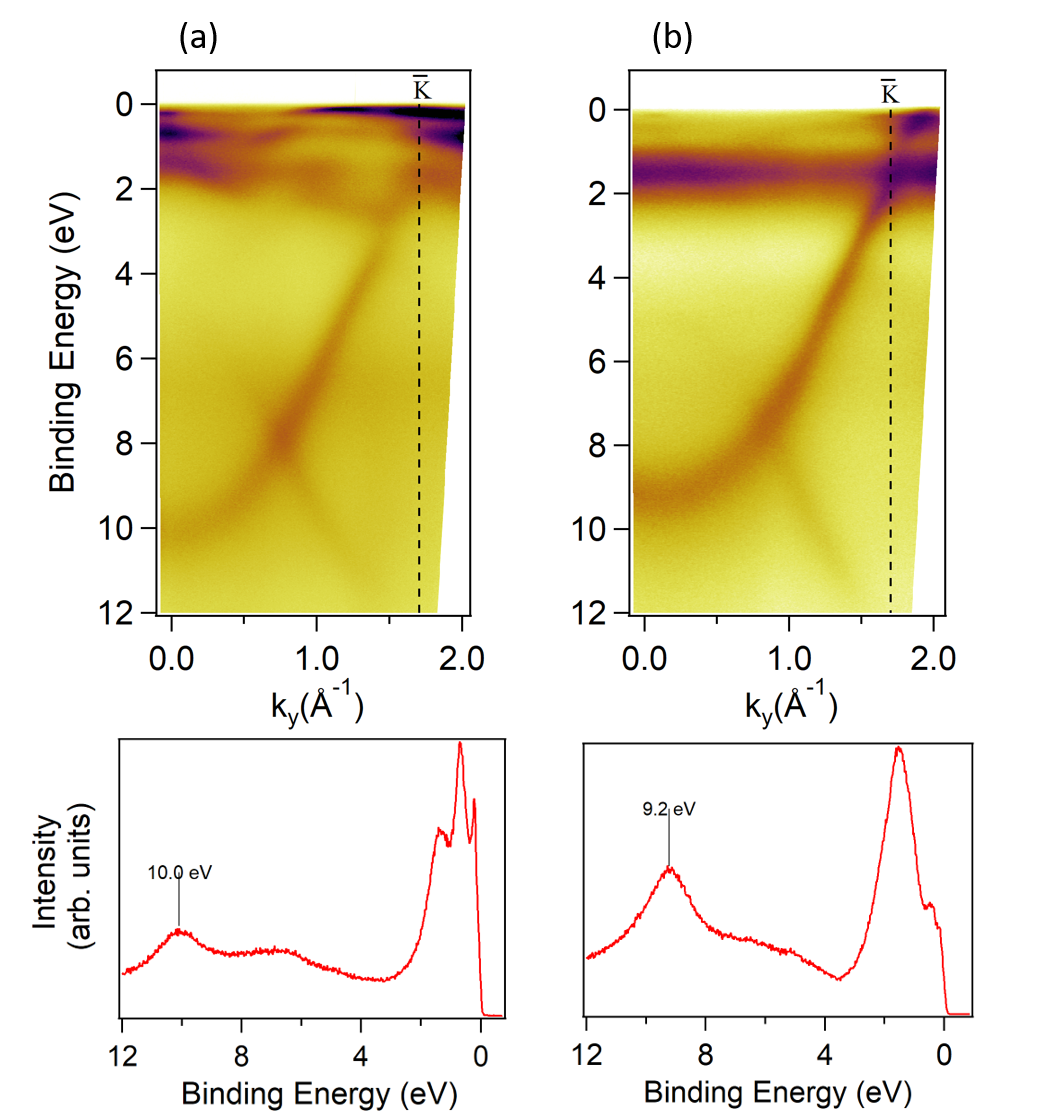}
\caption{\label{fig:befor after} 
 (a,b) ARPES maps along the $\bar\Gamma$-$\bar{\mathrm K}$ direction before and after Eu intercalation under graphene/Ni(111), respectively. Corresponding energy distribution curves on the lower part of the figure were obtained at the $\bar\Gamma$ point. The measurements were carried out at 70 eV photon energy at room temperature. In the graphene/Ni(111) (a), the $\pi$ band has a minimum at the $\bar\Gamma$ point at 10 eV of BE, and disperses towards the Fermi level up to about 2.3 eV, where interacts 
 with the Ni 3\textit{d} states around the $\bar{\rm K}$ point, in agreement with data reported in all previous studies. The Ni \textit{d}-states are weakly dispersing in the range 0-2 eV below the Fermi level over all the surface Brillouin zone. Upon the intercalation of Eu (b), the minimum of the $\pi$ band, at the $\bar\Gamma$ point, moves towards the Fermi level by about 0.8 eV, meaning a lower coupling with the substrate. The $\pi\pi^\ast$ band recovers the typical linear dispersion of quasi-freestanding graphene around the $\bar{\rm K}$ point below the Fermi level. 
}
\end{figure} 
\clearpage


\section{Spectral features due to the electron-phonon coupling}

In this Section of the Supplementary Materials we model
the effects of the retarded many-body
electron-phonon (el-ph) interaction in the ARPES spectra.
Such effects are evaluated by mapping the observed ARPES dispersion onto
a tight-binding (TB) model coupled with a phonon with typical energy $\omega_{\rm ph}=0.16$ eV. The strong anisotropy of the el-ph coupling caused by the closeness to the van Hove singularity (VHS) is modeled by means of different el-ph couplings along different directions.
We show that incoherent weakly-dispersive bands can be naturally accounted as a combined result of a sizable el-ph coupling
and of the proximity of the VHS to the Fermi level.
Similar polaronic features were reported in Refs. \cite{chen18,link19,chen23}

To our aim it is sufficient to consider a Holstein-like momentum-independent coupling of the electronic bands of graphene
with a dispersionless phonon mode with energy $\omega_0$.

The retarded electronic self-energy in the Matsubara space $\Sigma(i\omega_n)$ is computed as:
\begin{eqnarray}
\Sigma(i\omega_n)
&=&
\lambda T
\sum_m
\frac{\omega_0^2}{\omega_0^2+(\omega_n-\omega_m)^2}
G(i\omega_m)
,
\end{eqnarray}
where $\lambda$ is the dimensionless electron-phonon coupling constant,
$T$ the temperature, $i\omega_n$ and $i\omega_m$ fermionic Matsubara frequencies
and the average Green's function $G(i\omega_n)$ contains
the information about the band structure through
the integral over the electronic density of states $N(\epsilon)$:
\begin{eqnarray}
G(i\omega_n)
&=&
\int d\epsilon \frac{N(\epsilon)}{i\omega_n+\mu-\epsilon-\Sigma(i\omega_n)}
,
\label{gloc}
\end{eqnarray}
where $\mu$ is the chemical potential.
The electron-phonon self-energy $\Sigma(\omega)$ and the electronic Green's function $G(\omega)$
on the real-frequency axis are thus obtained by means of the standard Marsiglio-Schlossmann-Carbotte procedure \cite{msc},
and the electronic spectral function $A({\bf k},\omega)$, accessible through angle-resolved photoemission spectroscopy,
is computed as:
\begin{eqnarray}
A({\bf k},\omega)
&=&
-\frac{1}{\pi}
\mbox{Im}
\left[
\frac{1}{\omega-\epsilon_{\bf k}-\Sigma(\omega)}
\right]
.
\end{eqnarray}

In many wide-band conventional systems,
the effects of the self-energy in Eq. (\ref{gloc}) is safely neglected \cite{eliashberg, grimvall},
and the electron-phonon coupling gives rise to incoherent spectral features only at $\pm \omega_0$,
widely discussed in literature and responsible for the ``kink'' physics \cite{grimvall,scalaparks,lanzara,mazzola13}.
Such approximation fails however when relevant features
in the electronic density of states (band edges, van Hove singularities, \ldots)
lie close to the Fermi energy \cite{engelsberg,cp03,marsiglio05}.
In this situation the electron-phonon self-energy must be retained in
Eq. (\ref{gloc}), and it must be obtained
in a highly non-linear self-consistent way.
This non-linearity gives rise to multiphonon spectral features 
at energies $n \omega_0$ which are not expected
in the standard Eliashberg framework, leading eventually
to polaronic spectral profile.
The approach here employed is suitable in the low-intermediate coupling regime,
like pointed out in graphene, where the intensity of the
multiphonon spectral features $p(n\omega_0)$ follows a monotonic
Poisson distribution \cite{mahanbook}, whereas alternative methods must be employed
in the strong-coupling regime where the distribution $p(n\omega_0)$
smoothly evolves in a (non-monotopic) Gaussian one \cite{langfirsov,mahanbook,ciuchi97}.

\begin{figure}[t]
\includegraphics [width=0.9\textwidth ]{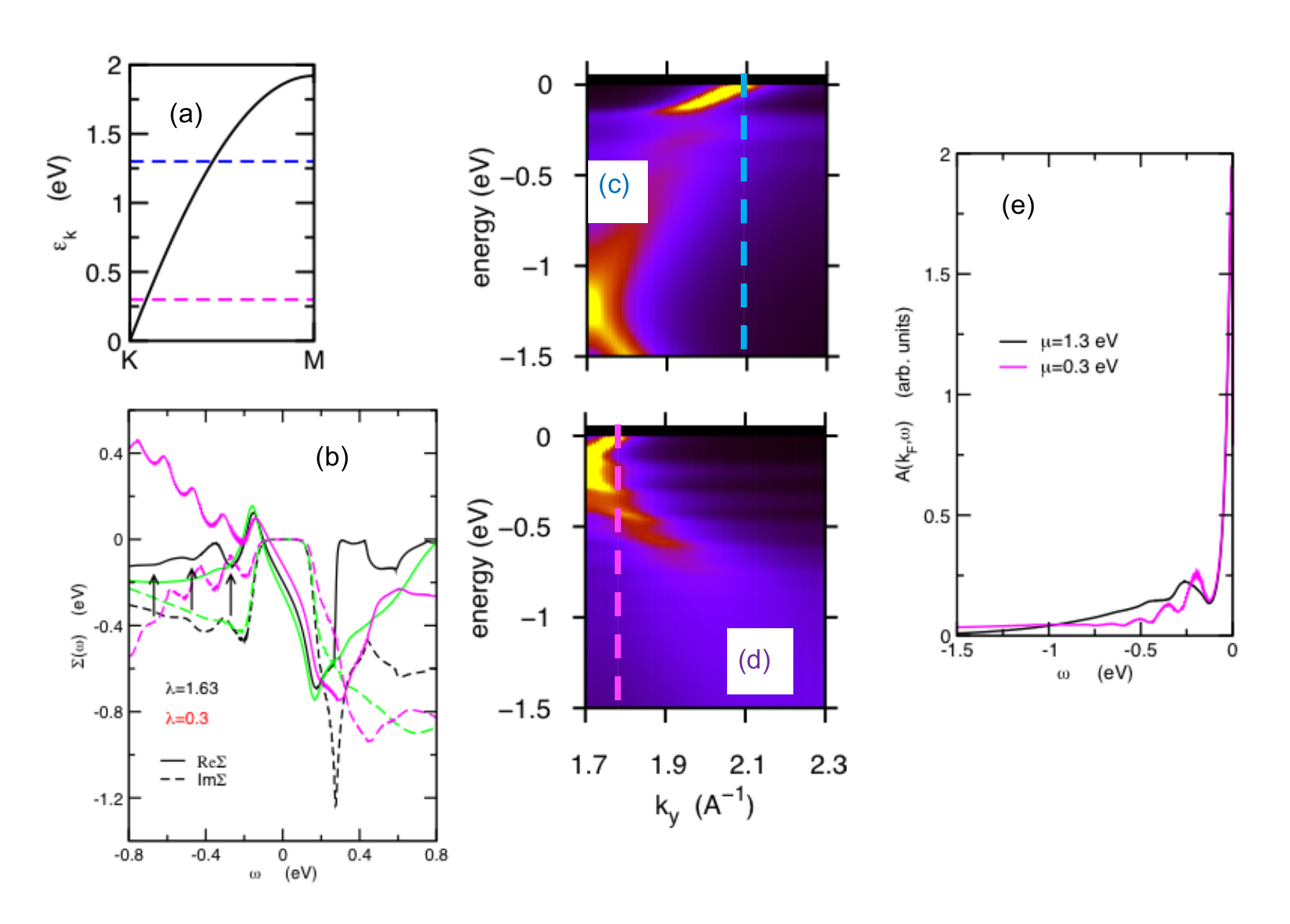}
\caption{(a) Electronic dispersion along the $\bar{\rm K}\bar{\rm M}$  cut. The relevant chemical potential
$\mu=1.3$ eV is drawn as blue dashed line. Also shown is the hypothetical case
of lower doped sample with $\mu=0.3$ eV (magenta dashed line).
(b) Frequency dependence of the real and imaginary parts of the electron-phonon self-energy for  $\mu=1.3$ eV
and an el-ph coupling constant $\lambda=1.63$, appropriate for the $\bar{\rm K}\bar{\rm M}$  cut (black lines).
The multiphonon features are marked by arrows.
Also shown are the self-energies for $\mu=1.3$ eV and $\lambda=0.3$
appropriate for the p($\bar{\Gamma}\bar{\rm K}$) cut (green lines),
and for $\mu=0.3$ eV,  $\lambda=1.63$ case (magenta lines). 
The self-energy for $\lambda=0.3$ has been magnified by a factor $1.63/0.3$ in order
to make it visible in the same scale.
In all the cases solid and dashed lines represent the real
and the imaginary parts of the self-energy, respectively. (c)-(d) color intensity maps for the cases $\mu=1.3$ eV and
$\mu=0.3$ eV. In both cases $\lambda=1.63$. (e) Comparison of the spectral function at $k_{\rm F}$ for 
panels (c) and (d), as marked by the light blue and magenta dashed lines in the corresponding panels.}
\label{f-elph}
\end{figure}

For the electronic structure, we employ the widely-used nearest-neighbor tight-binding model
where the effective interatomic hopping $t_0$ is chosen to best reproduce the experimental dispersion.
Focusing on the electron-phonon mediated features at the Fermi level, we disregard here the effects of the magnetic substrate
and we don't include a gap opening.
The chemical potential in set $\mu=1.3$ eV above the ungapped Dirac point
(here taken at zero energy as reference) as in ARPES data,
and we fix $t_0=1.92$ eV a Fermi momentum $k_{\bf F}\approx 2.1$ \AA$^{-1}$ along the
$\bar{\rm K}\bar{\rm M}$ direction $(0,k_y)$, as in Fig. 2b of the main text.
The van Hove singularity lies thus at the $\bar{\rm M}$ point at $0.62$ eV above the Fermi energy
(see Fig. \ref{f-elph}a).
This value probably overestimates the VHS distance from the Fermi level since
the interaction with substrate (see Fig. 1a) and correlation effects tends
to flatten the electronic dispersion close to the VHS saddle point \cite{link19}.

The presence of van Hove saddle point is expected
to yield a significant anisotropy in the electron-phonon coupling,
favouring small-${\bf q}$ scattering \cite{pattnaik92}.
The use of a fully momentum-resolved Eliashberg scheme is beyond
the purposes of the present analysis.
We simulated the anisotropy effects of the VHS by using
different electron-phonon coupling along different paths
of the Brillouin zone.
More in details, a dimensionless coupling $\lambda=1.63$
is estimated by the kink features of the ARPES data along the $\bar{\rm K}\bar{\rm M}$ cut,
crossing the VHS, whereas a mildest coupling $\lambda=0.3$ is observed
along the p($\bar{\Gamma}\bar{\rm K}$) cut, which lies far from the VHS.
The electron-phonon self-energy for the two cases is shown and compared in Fig. \ref{f-elph}b.
Most striking here is the appearance, for the strong-coupling case (black lines) along the $\bar{\rm K}\bar{\rm M}$ cut,
of multiphonon overtones, signalizing the emerging of polaronic features.
Such overtones are absent along the p($\bar{\Gamma}\bar{\rm K}$) cut with weaker coupling (green lines).
In order to highlight the role of the VHS in boosting polaronic features,
we consider also a template case of strong-coupling graphene,
with the same electron-phonon coupling $\lambda=1.63$, but
with smaller doping $\mu=0.3$ eV, so that the Fermi level
lies $1.62$ eV {\em below} the van Hove singularity.
The real part of the self-energy is shown in Fig. \ref{f-elph}b in magenta color.
Also in this case multiphonons overtones are visible, but with a smaller
imaginary part of the self-energy.
Panels (c) and (d) show the simulated ARPES intensity maps for both cases
($\mu=1.3$ eV and $\mu=0.3$ eV). Given the fixed equal electron-phonon coupling strength
($\lambda=1.63$) multiphonon overtones are visible in both cases.
It should be notes however that far from chemical potential far from the VHS ($\mu=0.3$ eV)
the multiphonon sidebands are quite limited in the momentum ${\bf k}$-space,
whereas they appears much spread over when the chemical potential approaches
the VHS ($\mu=1.3$ eV). In order to reveal the role of the VHS,
it is interesting also to analyze the profile of the spectral function $A({\bf k},\omega)$ for ${\bf k}={\bf k}_{\rm F}$
as accessible by angle-resolved spectroscopy.
For chemical potential far from the VHS ($\mu=0.3$ eV, magenta lines) the multiphonon features
are more visible, due to the smaller damping, but their intensity drastically reduces
with the energy.
On the other hand, when the Fermi level approaches the VHS ($\mu=1.3$ eV, black lines),
although the multiphonon peaks are less pronounced because of the large imaginary
part of the self-energy, the Poisson multiphonon distribution is clearly wider in energy,
signalizing a stronger polaronic feature.

\clearpage

\section{Berry curvature calculations for $(2\times2)$ graphene/Eu}

\begin{figure}[h]
\includegraphics[width=\textwidth]{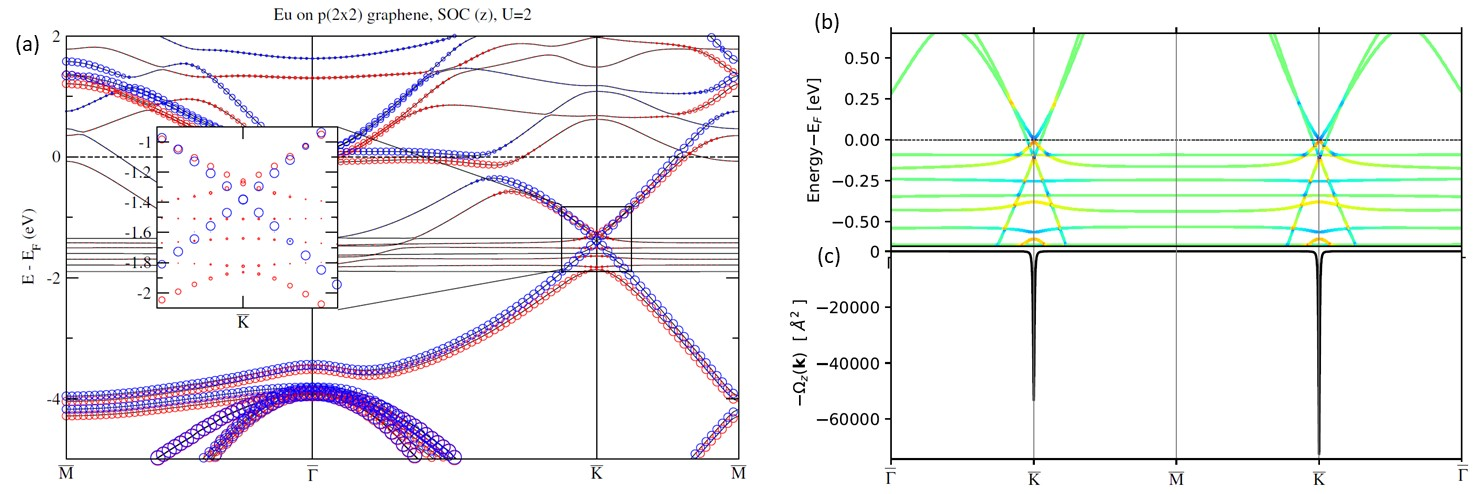}
\caption{\label{fig:BC} Berry curvature calculation. (a) Band structure of $(2\times2)$ reconstructed graphene on Eu (without Ni substrate) calculated using Wannier functions. A Hubbard correction U= 2.7 eV was used. The spin direction is out-of-plane. (b) \textit{k}-resolved Berry curvature for the system. The binding energy is referred to the Dirac point. The color scale is logarithmic with red=positive, green=zero, blue=negative. (c) Band-integrated \textit{k}-resolved Berry curvature, taken on the valence states. }

\end{figure} 


\end{document}